\documentclass[aps,prl,preprint,nopacs,superscriptaddress]{revtex4}
\usepackage{sidecap}
\usepackage{graphicx}
\usepackage{verbatim}
\usepackage{mathrsfs}
\usepackage{amsmath,amsfonts,amssymb}
\usepackage{graphicx}

\def\3{2.8in}    
\def\2{2.5in}
\def\4{3.0in}

\def \beq {\begin{equation}}
\def \eeq {\end{equation}}
\begin{document}

\title{Room-temperature magnetic topological Weyl fermion and nodal line semimetal states in half-metallic Heusler Co$_2$TiX (X=Si, Ge, or Sn)}
\author{Guoqing Chang$^*$}\affiliation{Centre for Advanced 2D Materials and Graphene Research Centre National University of Singapore, 6 Science Drive 2, Singapore 117546}\affiliation{Department of Physics, National University of Singapore, 2 Science Drive 3, Singapore 117542}
\author{Su-Yang Xu$^*$}\affiliation {Laboratory for Topological Quantum Matter and Spectroscopy (B7), Department of Physics, Princeton University, Princeton, New Jersey 08544, USA}
\author{Hao Zheng\footnote{These authors contributed equally to this work.}}\affiliation {Laboratory for Topological Quantum Matter and Spectroscopy (B7), Department of Physics, Princeton University, Princeton, New Jersey 08544, USA}

\author{Bahadur Singh}
\affiliation{Centre for Advanced 2D Materials and Graphene Research Centre National University of Singapore, 6 Science Drive 2, Singapore 117546}
\affiliation{Department of Physics, National University of Singapore, 2 Science Drive 3, Singapore 117542}
\author{Chuang-Han Hsu}\affiliation{Centre for Advanced 2D Materials and Graphene Research Centre National University of Singapore, 6 Science Drive 2, Singapore 117546}
\affiliation{Department of Physics, National University of Singapore, 2 Science Drive 3, Singapore 117542}

\author{Guang Bian}\affiliation{Laboratory for Topological Quantum Matter and Spectroscopy (B7), Department of Physics, Princeton University, Princeton, New Jersey 08544, USA}
\author{Nasser Alidoust}\affiliation {Laboratory for Topological Quantum Matter and Spectroscopy (B7), Department of Physics, Princeton University, Princeton, New Jersey 08544, USA}

\author{Ilya Belopolski}\affiliation {Laboratory for Topological Quantum Matter and Spectroscopy (B7), Department of Physics, Princeton University, Princeton, New Jersey 08544, USA}
\author{Daniel S. Sanchez}\affiliation {Laboratory for Topological Quantum Matter and Spectroscopy (B7), Department of Physics, Princeton University, Princeton, New Jersey 08544, USA}

\author{Hsin Lin$^{\dag}$}
\affiliation{Centre for Advanced 2D Materials and Graphene Research Centre National University of Singapore, 6 Science Drive 2, Singapore 117546}
\affiliation{Department of Physics, National University of Singapore, 2 Science Drive 3, Singapore 117542}

\author{M. Zahid Hasan\footnote{Corresponding authors: nilnish@gmail.com, mzhasan@princeton.edu}}\affiliation {Laboratory for Topological Quantum Matter and Spectroscopy (B7), Department of Physics, Princeton University, Princeton, New Jersey 08544, USA}

\begin{abstract}
Topological semimetals (TSMs) including Weyl semimetals and nodal-line semimetals are expected to open the next frontier of condensed matter and materials science. Although the first inversion breaking Weyl semimetal was recently discovered in TaAs, its magnetic counterparts, i.e., the time-reversal breaking Weyl and nodal line semimetals, remain elusive. They are predicted to exhibit exotic properties distinct from the inversion breaking TSMs including TaAs. In this paper, we identify the magnetic topological semimetal state in the ferromagnetic half-metal compounds Co$_2$TiX (X=Si, Ge, or Sn) with Curie temperatures higher than 350 K. Our first-principles band structure calculations show that, in the absence of spin-orbit coupling, Co$_2$TiX features three topological nodal lines. The inclusion of spin-orbit coupling gives rise to Weyl nodes, whose momentum space locations can be controlled as a function of the magnetization direction. Our results not only open the door for the experimental realization of topological semimetal states in magnetic materials at room temperatures, but also suggest potential applications such as unusual anomalous Hall effects in engineered monolayers of the Co$_2$TiX compounds at high temperatures.
\end{abstract}

\date{\today}
\maketitle
Topological semimetals (TSM) are electronic strong spin-orbit metals/semimetals whose Fermi surfaces arise from crossings between conduction and valence bands, which cannot be avoided due to nontrivial topology \cite{rev1, TI_book_2014_2, TI_book_2015}. Such new states of topological matter have recently attracted worldwide interest because they may realize particles that remain elusive in high energy physics, exhibit quantum anomalies, host new topological surface states such as the Fermi arc and the drumhead surface states, and show exotic transport and spectroscopic behaviors arising from the novel bulk and surface topological band structures \cite{Wan, Hasan_Na3Bi, Weyl1, Weyl2, NL1, NL2,CA1, CA2, TaAs2, TaAs1,CA3, CA4, STM}. Among the proposed TSM states, two of the most exciting ones are the Weyl semimetal \cite{Weyl1, Weyl2, Weyl3, Weyl4, Weyl5, Weyl6} and the nodal line semimetal states \cite{CaP, CuPdN, LaX, Guang1, Guang2}. While the Fermi surface of a Weyl semimetal consists of isolated 0D points in $k$ space, i.e., the Weyl nodes, the Fermi surface of a nodal-line semimetal is a 1D closed loop, i.e., the nodal line winding in 3D momentum space. Although the first inversion breaking Weyl semimetal was recently discovered in TaAs\cite{Weyl1, Weyl2, Weyl3}, the time-reversal breaking Weyl and nodal line semimetals remain elusive. The time-reversal breaking (magnetic) Weyl and nodal line TSM states are predicted to show exotic properties beyond the inversion-breaking TSMs such as TaAs. Firstly, ferromagnetic materials usually have considerable electron-electron interaction. Hence, a magnetic TSM is a promising platform to study the interplay between the TSM state and electronic correlation, which may potentially lead to new correlated topological phases \cite{Wan, Mott1, Mott2}. Secondly, magnetic Weyl semimetals show the anomalous Hall effect \cite{QAH, AH}, i.e., Hall-like conductivity without an external magnetic field. When making such a magnetic Weyl semiemtal into 2D (monolayer), the anomalous Hall conductance may be quantized. To date, the quantum anomalous Hall effect has only been observed in magnetically doped topological insulator thin film samples such as Cr$_x$(Bi$_y$Sb$_{1-y}$)$_{2-x}$Te$_3$ \cite{Xue}, which required ultra-low (mK) temperatures. By contrast, monolayer samples of magnetic Weyl semiemtals, which are natural ferromagnets, may realize the quantum anomalous Hall effect at significantly higher temperatures, and therefore make this novel phenomena relevant in actual applications \cite{QAH}. Thirdly, the superconducting proximity effect of a magnetic Weyl semimetal is predicted to show topological Weyl superconductivity \cite{Weyl-SC}. In such an exotic topological superconductor, the superconducting gap has point nodes, which are Weyl nodes, and the Weyl nodes are connected by Majorana Fermi arcs on the surface.  

Despite interest, to date, magnetic topological semimetals remain experimentally elusive. One main difficulty is that ferromagnetic semimetals, regardless of its topological trivial/nontrivial nature, are rare in nature. The existing proposed magnetic TSM materials, such as the pyrochlore iridates and HgCr$_2$Se$_4$ \cite{Wan, QAH} have relatively low magnetic transition temperatures that are much lower than the room temperature. This fact not only hinders the experimental confirmation but also makes the predicted TSM state inaccessible in actual applications. Here, we present our identification of the room-temperature topological Weyl and nodal line semimetal states in half-metallic Co$_2$TiX. 

A half-metal is a type of ferromagnet that acts as a conductor to electrons of one spin, but as an insulator or semiconductor to those of the opposite spin. The half metallicity suggests it as a promising candidate for the TSM state because both the Weyl semimetal and nodal-line semimetal states require crossings between two singly degenerate (spin polarized) bands.In this paper, we explore the possible existence of the TSM state in the half-metallic full Heusler compounds. The half-metallicity in full Heusler compounds including Co$_2$MnSi,  Co$_2$MnGe, Co$_2$FeAl$_{0.5}$Si$_{0.5}$, and Co$_2$TiX (X=Si, Ge, or Sn) has been a well-known phenomenon in both theory and experiments  \cite{CoTiSi1, CoTiSi2, C2, CoMnSi, C3, C18, C19, C20, C21, C22, C25, C26, C27}. We focus on the Co$_2$TiX (X=Si, Ge, or Sn).

Electronic band structures were calculated within the density functional theory (DFT) \cite{DFT1} framework with the projector augmented wave (PAW) method, using the VASP (Vienna Ab Initio Simulation Package) \cite{DFT2, DFT3}. The generalized gradient approximation was used to describe the exchange-correlation effects \cite{DFT4}. We used a kinetic energy cut-off of 500 eV and a 16$\times$16$\times$16-centered $k$-mesh to sample the primitive bulk Brillouin zone (BZ). In order to compute the bulk band structures, we used the experimental lattice constants \cite{CoTiSi1}, $a=5.770$ $\textrm{\AA}$, $a=5.830$ $\textrm{\AA}$, and 5.997 $\textrm{\AA}$ for Co$_2$TiSi, Co$_2$TiGe and Co$_2$TiSn, respectively. The spin-orbit coupling was employed in the electronic structure calculations as implemented in the VASP.

\begin{figure*}
\centering
\includegraphics[width=13cm]{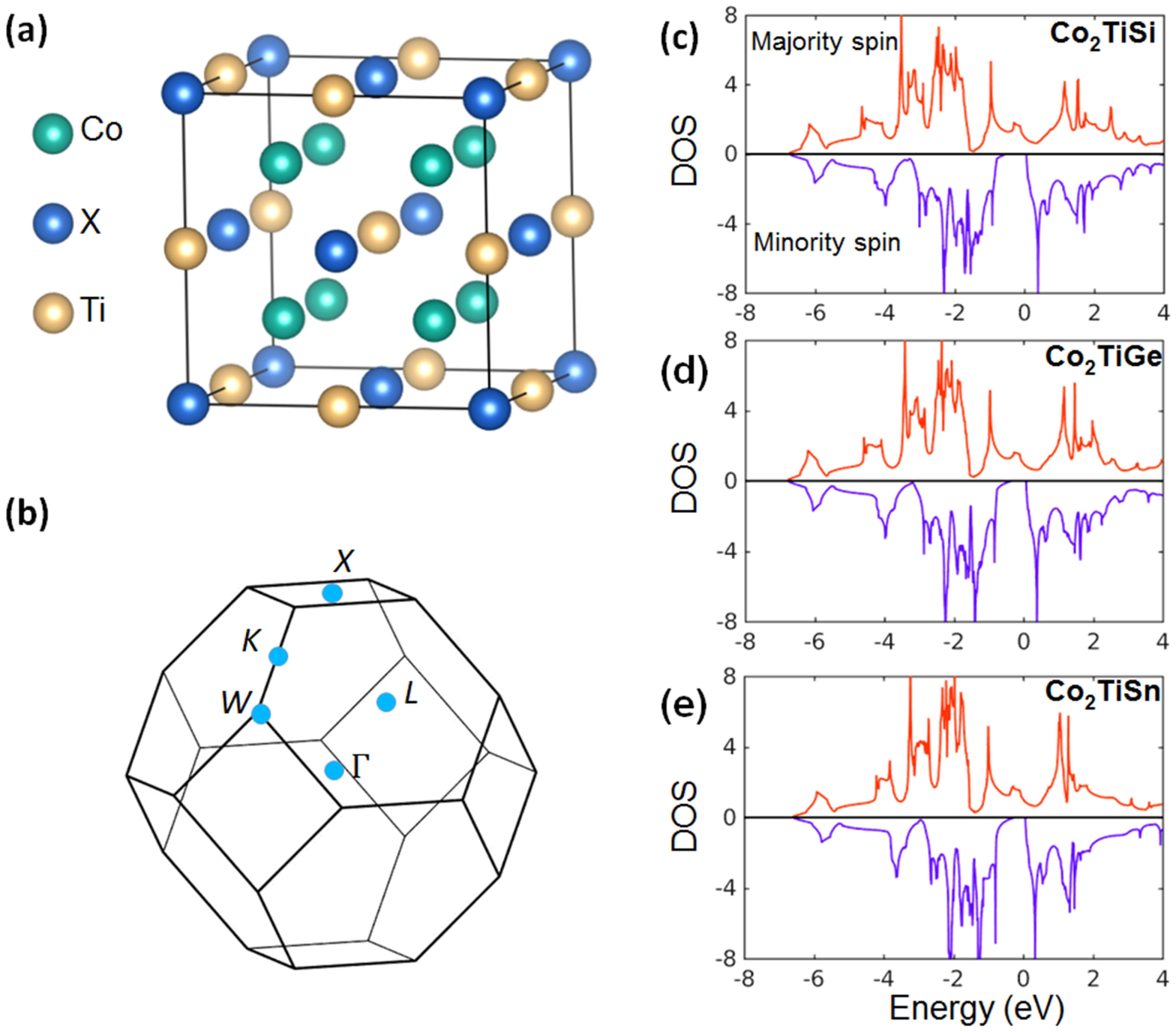}
\caption{\label{Fig1}\textbf{Crystal structure and density of states of the Co$_2$TiX compounds.}
(a) The face-centered cubic structure of the Co$_2$TiX Heusler compounds. The Co, Ti and X (X=Si, Ge, Sn) atoms are represented by the green, yellow, and blue balls, respectively. (b) The first bulk Brillouin zone. High symmetry points are marked. (c-e) Calculated spin-resolved density of states (DOS) of Co$_2$TiX without spin-orbit coupling.}
\end{figure*}

Co$_2$TiX crystalizes in a face-centered cubic (FCC) lattice with the space group Fm$\bar{3}$m (Fig.~\ref{Fig1}(a)). Previous magnetic (SQUID) measurements clearly established the ferromagnetic groundstate in these compounds. The relevant symmetries are the 3 mirror planes, $\mathcal{M}_x$ $(k_x=0)$, $\mathcal{M}_y$ $(k_y=0)$, $\mathcal{M}_z$ $(k_z=0)$, and three C4 rotation axes, kx, ky and kz.The Curie temperatures are 380 K for Co$_2$TiSi, and Co$_2$TiGe and 355 K for Co$_2$TiSn. Figure~\ref{Fig1}(c-e) show the calculated spin-resolved density of states (DOS). We clearly see that, for all three compounds, the band structure is fully gapped at the Fermi level for one spin, i.e., the minority spin, whereas it is metallic for the other spin, i.e., the majority spin. This demonstrates the half-metallic groundstate of Co$_2$TiX, consistent with previous theoretical and experimental studies \cite{CoTiSi1, CoTiSi2, C19, C22}. 

\begin{figure*}
\centering
\includegraphics[width=17cm]{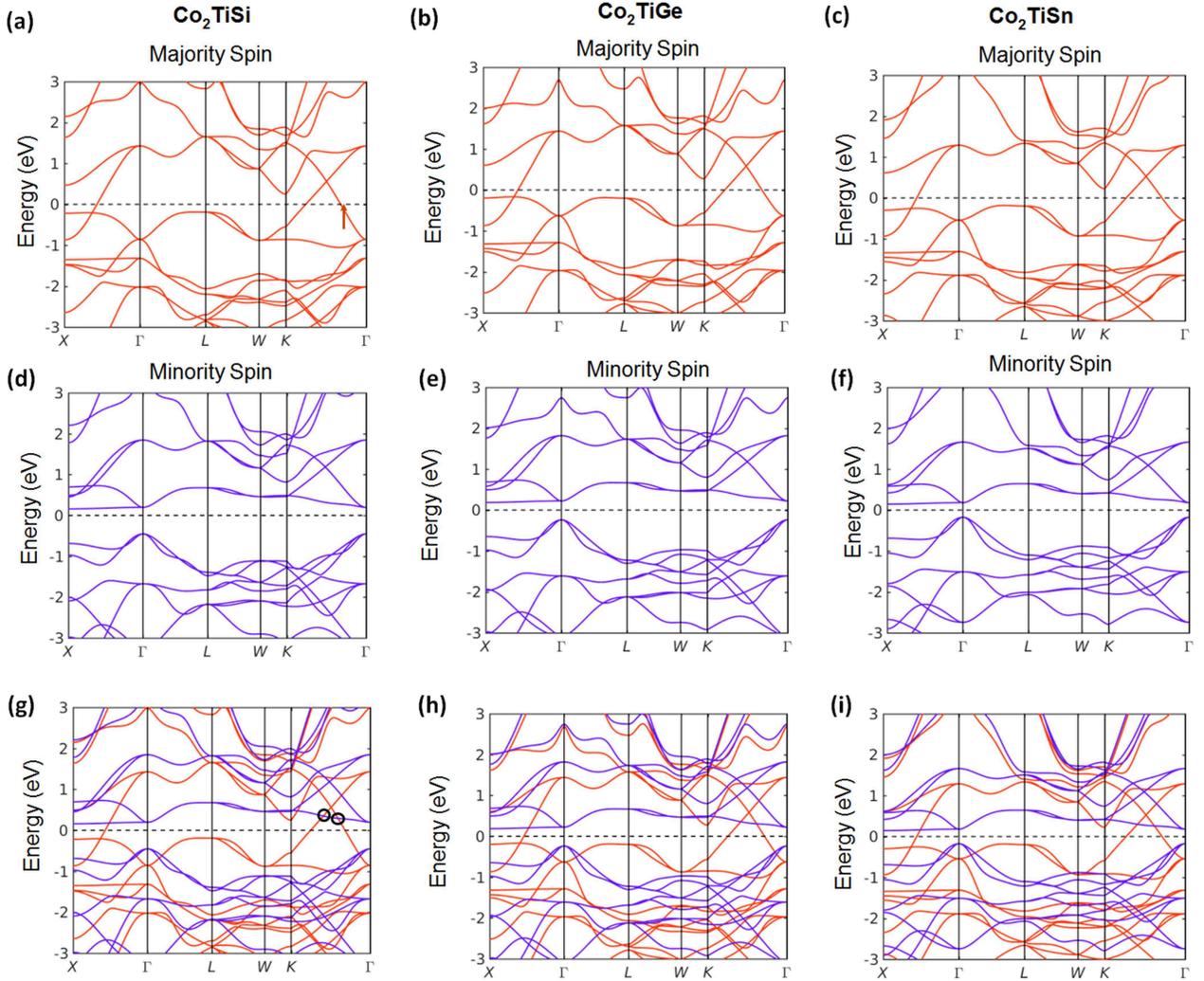}
\caption{\label{Fig2}\textbf{Spin-resolved band structure and ferromagnetic half-metallic ground states in Co$_2$TiX.}
(a-c) The calculated bulk band structures of the majority spin of the Co$_2$TiSi, Co$_2$TiGe, and Co$_2$TiSn, respectively. (d-f) Same as panels (a-c) but for the minority spin. (g-i) The band structures of both spins.}
\end{figure*}

Figure~\ref{Fig2} shows the first-principles calculated band structure along high-symmetry lines in the absence of spin-orbit coupling. For the band structures of the minority spin (Figs.~\ref{Fig2}(a-c)), we obtain a minority spin band gap of about 0.5 eV, which is in agreement with previous spin-resolved x-ray absorption spectroscopic measurements \cite{CoTiSi2}. On the other hand, the majority spin band structures (Figs.~\ref{Fig2}(d-f)) show clear band crossings between the conduction and valence bands along the $\Gamma-X$ and $\Gamma-K$ directions. Furthermore, additional band crossings between the majority and minority spins (highlighted by the black circles in Figs.~\ref{Fig2}(g-i)) are identified as we overlay the band structures of the two spins.

\begin{figure*}
\centering
\includegraphics[width=13cm]{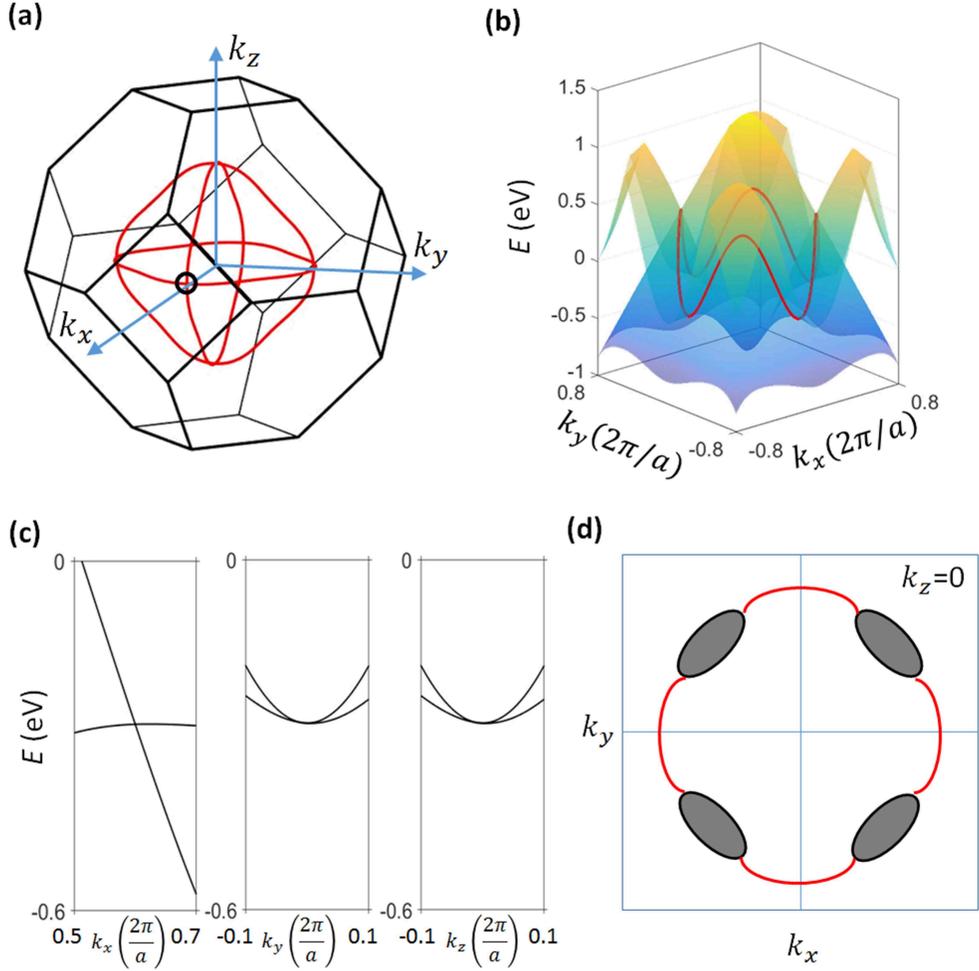}
\caption{\label{Fig3}\textbf{The topological nodal-lines in Co$_2$TiX in the absence spin-orbit coupling.}
(a) The nodal-lines in $k_x$-$k_y$-$k_z$ space formed by the band crossings within the majority band structure. (b) The band structure in $E$-$k_x$-$k_y$ space containing the nodal-line on the $k_z=0$ plane. On $k_z$=0 plane, the bulk valence and conduction bands dip into with each other and form a energy-dependent nodal-line (the red line). (c) The energy dispersions along $k_x$ (left panel), $k_y$ (middle panel), and $k_z$ (left panel) directions away from the crossing point surrounded by the black circle in panel (a). (d) Band crossings on the $k_z=0$ plane. The red lines show the nodal line formed by the same (majority) spin. The Black ellipses show the cross-sections of the nodal-surfaces formed by opposite spins.}
\end{figure*}

In order to understand the momentum space configuration of the band crossings without spin-orbit coupling, we calculate the band structure at all $k$ points throughout the bulk BZ. Figure~\ref{Fig3}(a) shows the band crossings in the first BZ. Specifically, we find that the crossings within the majority spin band structure form three nodal lines around the $\Gamma$ point on the $k_x=0$, $k_y=0$, and $k_z=0$ planes. Figures~\ref{Fig3}(c) show the energy dispersions along $k_x$, $k_y$, and $k_z$ directions that cut across a $k$ point on the $k_z=0$ nodal line as noted by the black dot in Fig.~\ref{Fig3}(a). We see that the two bands disperse linearly away from the crossing point along the radial ($k_x$) direction. On the other hand, the dispersion becomes quadratic along the tangential direction. These dispersions confirm the existence of nodal lines. The two crossing bands have opposite mirror eigenvalues, confirming these nodal rings in the majority spin channels are protected by mirror symmetry. We now consider the band crossings between bands of opposite spins. Since the bands of opposite spins do not hybridize without spin-orbit coupling, the crossings form several 2D closed surfaces, i.e., nodal-surfaces, in the BZ. The situation on the $k_z=0$ plane is shown in Fig.~\ref{Fig3}(d). In addition to the nodal-line (the red lines), four nodal-surfaces also cross this plane inside the four quadrants.

\begin{figure*}
\centering
\includegraphics[width=13cm]{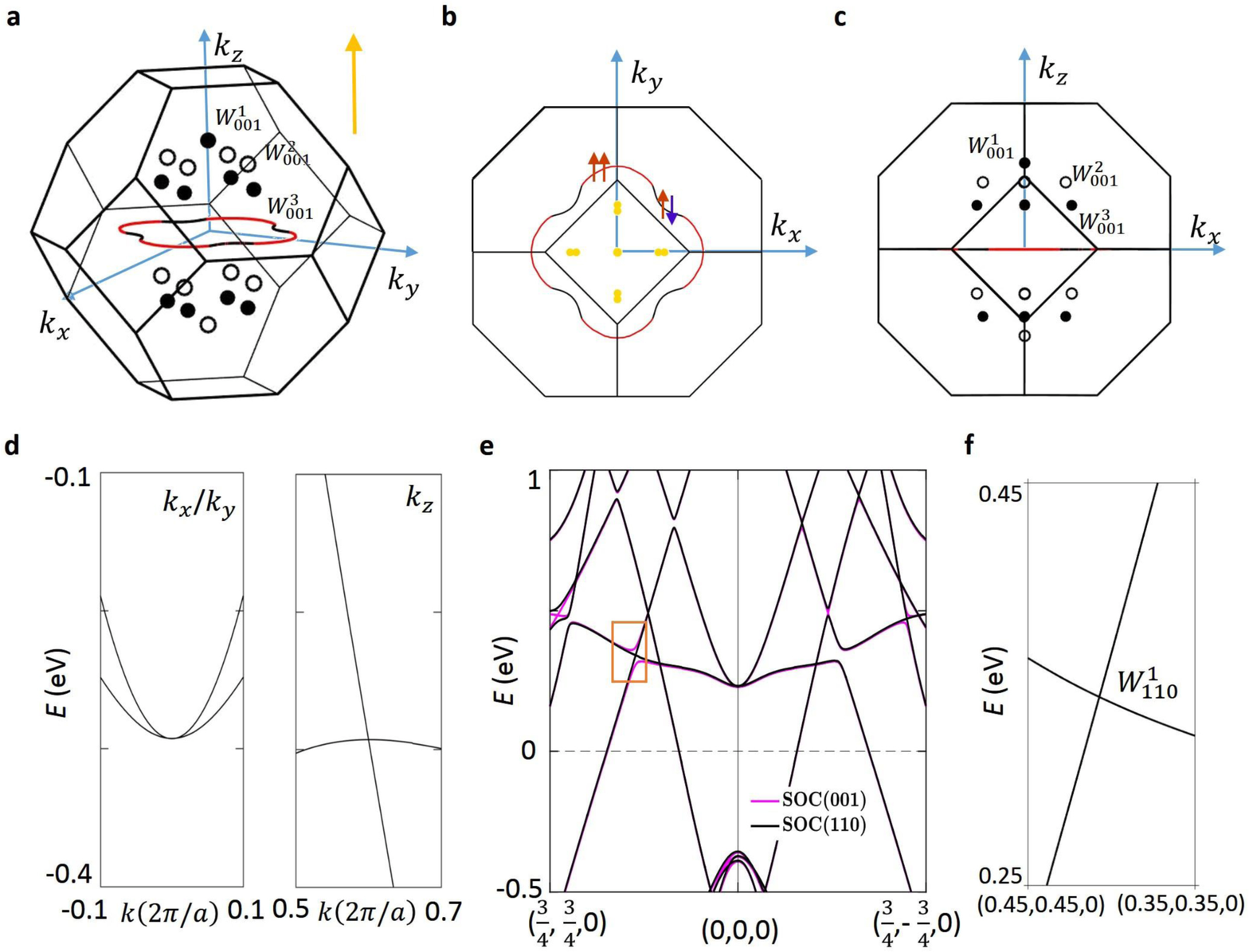}
\caption{\label{Fig4}\textbf{Weyl nodes and nodal-lines in Co$_2$TiX in the presence of spin-orbit coupling. }
(a) Band crossings in the first BZ. There are one nodal line on $k_z$=0 plane and three types of Weyl nodes (shown by the dots) in the first BZ. The white and black colors denote the chiral charge of the Weyl nodes. (b) The projection of the band crossings on the (001) top surface. The projected Weyl nodes are shown by the yellow dots, all of which have a projected chiral charge of 0 as two Weyl nodes of opposite chiralities are projected onto the same point on the (001) surface. The nodal line in panels (a,b) is shown by the solid lines which are in two colors (red and black). The red segments denote the band crossings formed by two bands of the same (majority) spin whereas the black ones are formed by bands of opposite spins. (c) The projection of the band crossings on the (010) side surface. (d) The energy dispersions along $k_x$/$k_y$ (left panel) and $k_z$ (right panel) directions of $W_{001}^1$ Weyl cone.  The quadratic touching of two bands along $k_x$/$k_y$ direction proves that the chiral charge of the Weyl cone is $\pm$2. (e) A comparison of the band structures of Co$_2$TiX with a (001) or (110) magnetization direction. The number and momentum space locations of the Weyl nodes critically depend on the magnetization directions. (f) The zoom-in view of the area indicated by the orange box in panel (e). Unlike the case in the (001) magnetization, the band crossing remains intact with the (110) magnetization, and therefore becomes a Weyl node, the $W_{110}^1$. }
\end{figure*}

In Figure~\ref{Fig4}, we show the band structure after the inclusion of spin-orbit coupling. In the presence of spin-orbit coupling, the symmetry of the system and the electronic structures depend on the magnetization direction. We have calculated the free energy of the system with the magnetization direction along the (001), (110), and (111) directions. Our results show that the difference of the free energy along different magnetization directions is below 0.1 meV, which is beyond the resolution of DFT, suggesting that the system's magnetization direction can be controlled by an external magnetic field. We present systematic calculation results with the (001) magnetization. With the magnetization along the (001) direction and in the presence of spin-orbit coupling, only the $\mathcal{M}_z$ mirror symmetry and the $C_{4z}$ rotational symmetry are preserved. Hence we expect only the nodal line on the $k_z=0$ plane to survive. Indeed, we found that this is the case as shown in Figs.~\ref{Fig4}(a,b). Near the $k_x$ and $k_y$ axes, the nodal line is formed by bands of the same (majority) spin and is indicted by red color. By contrast, along the $k_x=k_y$ and $k_x=-k_y$ axes ($45^{\circ}$), the crossing happens between opposite spins and is denoted by black color. Since spin is not a good quantum number in the presence of spin-orbit coupling, hence the above description is approximate. On the other hand, the other two nodal-lines on the $k_x=0$ and $k_y=0$ planes are gapped out as the respective mirror symmetries are broken by the inclusion of spin-orbit coupling and the (001) magnetization. As a result, we expect Weyl nodes to emerge. Specifically, we find three-types of Weyl nodes, noted as $W_{001}^1$, $W_{001}^2$, and $W_{001}^3$ in Figs.~\ref{Fig4}(a-c), respectively, where the subscript denotes the magnetization direction. The $W_{001}^1$ are located on the $k_z$ axis and they are quadratic double Weyl nodes with chiral charge of $\pm2$. The $W_{001}^2$, and $W_{001}^3$ are at arbitrary $k$ points in the BZ and are single Weyl nodes with chiral charge of $\pm1$. The energy and momentum space locations of the Weyl nodes are shown in Table~\ref{Weyl}. Fig.~\ref{Fig4}(d) shows the energy dispersion away from the $W_{001}^1$ Weyl node. The quadratic $W_{001}^1$ Weyl nodes are protected by the $C_{4z}$ rotational symmetry. Indeed, we see that the two bands disperse linearly along the $k_z$ direction but quadratically along the $k_x,k_y$ directions. Finally, we briefly discuss the band structure with a (110) magnetization as the groundstate. Fig.~\ref{Fig4}(e) shows a comparison between band structures with either a (001) or (110) magnetization direction. It can be seen that the two band structures are very similar, which is consistent with our conclusion that the free energy values with different magnetization directions are quite close to each other. We highlight the area enclosed by the orange box, which is the band structure along the (110) (the $k_x-k_y=0$ and $k_z=0$) direction. While the band structure with (001) magnetization shows an avoided crossing, i.e., a gap, inside the orange box, the band structure with (110) magnetization shows a band crossing. In this case, the three mirror symmetries ($\mathcal{M}_x$, $\mathcal{M}_y$ and $\mathcal{M}_z$) and the three $C_4$ rotational symmetries ($C_{4x}$, $C_{4y}$, and $C_{4z}$) are broken. The preserved symmetries are the Mirror symmetry $\mathcal{M}_{xy}$ (the mirror plane normal to the (110) direction), the $C_2$ rotational symmetry along the (110) direction, and the inversion symmetry. Therefore all the nodal rings will gap out. We note that the band crossing along (110) direction is protected as the two bands have the opposite eigenvalues of the $C_2$ rotational symmetry. We have checked the chiral charge by calculating the Berry flux through a 2D closed surface in $k$ space enclosing this crossing point and found that it is indeed a Weyl node. We note that this Weyl node, $W_{110}^1$, does not exist in the band structure with the (001) magnetization but arise with the (110) magnetization. This fact suggests a novel possibility that the number, the momentum space location, and other properties of the Weyl nodes in the Co$_2$TiX system can be engineered by tuning the magnetization direction.

We discuss the effect of the onsite Coulomb repulsion $U$ to the band structure and the Weyl nodes. We use $U=0$ in our paper.  First, we emphasize that a previous work \cite{CoTiSi1} on Co$_2$TiX that combined both experimental measurements and first-principles calculations showed that the calculated band structure with $U=0$ actually fits the experimental results better. Therefore, we believe that $U=0$ better reflects the experimental reality. Nevertheless, we have also calculated the band structure with a finite $U$ value. We found that a finite $U$ does not have a significant effect on the majority spin band structure. On the other hand, a finite $U$ does increase the band gap of the minority spin. Using this as a guideline, we can understand how a finite $U$ value affects the Weyl nodes in our calculations. We take the (001) magnetization as an example. As we have discussed in Fig.~\ref{Fig4}(a), we find three-types of Weyl nodes, noted as $W_{001}^1$, $W_{001}^2$, and $W_{001}^3$. We further note that $W_{001}^1$ arises from the crossing between two mainly majority spin bands, whereas $W_{001}^2$ and $W_{001}^3$ arise from the crossing between one majority spin and one minority spin bands. Therefore, upon the inclusion of a finite $U$ value, $W_{001}^1$ is hardly affected, while $W_{001}^2$ and $W_{001}^3$ are pushed to higher energies away from the Fermi level. We emphasize that the above description is approximate because spin is not a good quantum number in the presence of spin-orbit coupling. However, considering the fact that spin-orbit coupling is not very strong in Co$_2$TiX (X=Si, Ge, or Sn). The mixing between majority and minority spins is not expected to be significant.

In conclusion, we have identified the magnetic topological Weyl and nodal line semimetal states in the ferromagnetic half-metal compounds Co$_2$TiX (X=Si, Ge, or Sn) with Curie temperatures higher than 350 K. Our results pave the way for realizing topologically protected emergent properties in magnetic semimetals at room temperatures, highlighting the potential for electronics and spintronics applications in the Co$_2$TiX-based compounds.

\clearpage
\clearpage

\clearpage
\begin{center}
\begin{table}
\begin{tabular}{p{3cm}p{2cm}p{3cm}p{3cm}p{2cm}p{2cm}}
\hline
Weyl nodes & $k_x$ ($\frac{\pi}{a}$) & $k_y$ ($\frac{\pi}{a}$) & $k_z$ ($\frac{\pi}{a}$) & Charge & $E$ (eV)\\
\hline
$W_{001}^1$ & 0.00 & 0.00 & 0.60 & -2 & -0.285 \\
$W_{001}^2$ & 0.00 & -0.29 & 0.46 & +1 & 0.315 \\
$W_{001}^3$ & 0.00 & -0.33 & 0.30 & -1 & 0.315 \\
\hline
\end{tabular}
\caption{\label{Weyl} \textbf{Energy and momentum space locations of the Weyl nodes in Co$_2$TiGe with a (001) magnetization.}}

\end{table}
\end{center}

\end{document}